# Leveraging Retrieval Augmented Generative LLMs for Automated Metadata Description Generation to Enhance Data Catalogs

Mayank Singh[1], Abhijeet Kumar[1], Sasidhar Donaparthi[1], Gayatri Karambelkar[1]

[1]Fidelity Investments, Bangalore, Karnataka

mayank.singh@fmr.com  abhijeet.kumar@fmr.com  sasidhar.donaparthi@fmr.com
gayatri.karambelkar@fmr.com

## ABSTRACT

*Data catalogs serve as repositories for organizing and accessing diverse collection of data assets, but their effectiveness hinges on the ease with which business users can look-up relevant content. Unfortunately, many data catalogs within organizations suffer from limited searchability due to inadequate metadata like asset descriptions. Hence, there is a need of content generation solution to enrich and curate metadata in a scalable way.*

*This paper explores the challenges associated with metadata creation and proposes a unique prompt enrichment idea of leveraging existing metadata content using retrieval based fewshot technique tied with generative large language models (LLM). The literature also considers finetuning an LLM on existing content and studies the behavior of few-shot pretrained LLM (Llama, GPT3.5) vis-à-vis few-shot finetuned LLM (Llama2-7b) by evaluating their performance based on accuracy, factual grounding, and toxicity. Our preliminary results exhibit more than 80% Rouge-1 F1 for the generated content. This implied 87%-88% of instances accepted as is or curated with minor edits by data stewards. By automatically generating descriptions for tables and columns in most accurate way, the research attempts to provide an overall framework for enterprises to effectively scale metadata curation and enrich its data catalog thereby vastly improving the data catalog searchability and overall usability.*

## KEYWORDS

*Content Generation, NLG, Generative LLMs, Few-Shot Prompting, Data Catalog, Metadata Enrichment*

## 1. INTRODUCTION

In the modern digital ecosystem, locating relevant data has become increasingly challenging due to the rapid expansion of data assets. A Data Catalog combines metadata with data management and search tools, enabling efficient organization and access to vast amounts of information. These catalogs play a pivotal role in promoting data discovery, governance, and collaboration [1], assisting users in finding the data and solving redundancy within an organization [2]. According to a study by Gartner [3] the metadata management software market grew at 21.6%, reaching $1.54 billion in U.S. dollars. These solutions, designed to manage datasets, heavily rely on metadata that describes the data sources. However, traditional data catalogs and governance methodologies typically rely on data teams to do the heavy lifting of manual data entry, holding them responsible for updating the catalog as data assets evolve. This approach is not only time-intensive but requires significant manual toil. All these barriers result in limiting the availability of metadata. [4] Many organizations are currently transitioning legacy databases to data catalogs,

which often lack high-quality metadata. This metadata primarily covers physical dataset attributes, historical access, ownership etc. and often contains only partial descriptions of tables and columns. It fails to provide context regarding the actual content within the data or guidance on utilizing multiple datasets effectively. [5][6]

In this study, we explore the application of generative AI techniques to automate metadata generation for data catalogs, specifically the table and column descriptions. We leverage existing curated information from a data catalog and systematically incorporate it into large language models (LLMs) as illustrated in Figure 1. The fields present in the cyan box to the left shown in the Figure 1 are the asset information we have at start, which are then used to fetch more relevant information through intermediate steps that enrich the prompt going to the LLM. By harnessing the contextual generation capabilities of LLMs, we establish a robust foundation for data stewards, who can then refine and enhance these generated descriptions, significantly reducing the time required for manual composition and review. This use-case confines to an open text generation problem where it is hard to assess factual grounding, hallucination, or contradictions unless the ground truth is available (refer Section 4.2 and 4.3 for evaluation approaches).

The paper proposes an overall design of LLM-based framework effective for metadata content generation using relevant context in data catalog. Major contribution presented are:

1. A uniquely crafted semantic retrieval & longest common sequence (LCS) based re-ranking process for fetching similar column for few-shot prompting (way to bring domain knowledge to model).

2. A carefully mapped expander module for abbreviated column and table names (truly relevant in enterprise data catalog) for enriching prompt and assisting LLM to generate accurate descriptions.

3. A thorough study of the model performance and behavior of few-shot pretrained LLM and domain finetuned LLM setup. Broader feedback and experimental results on proprietary dataset establish efficacy of developing such system.

The remainder of this paper is organized as follows: Section 2 discusses related work in the field, highlighting existing approaches and gaps. Section 3 details our methodology for generative AI-based metadata creation. In Section 4, we present our results and discuss their implications. Finally, Section 5 concludes the study.

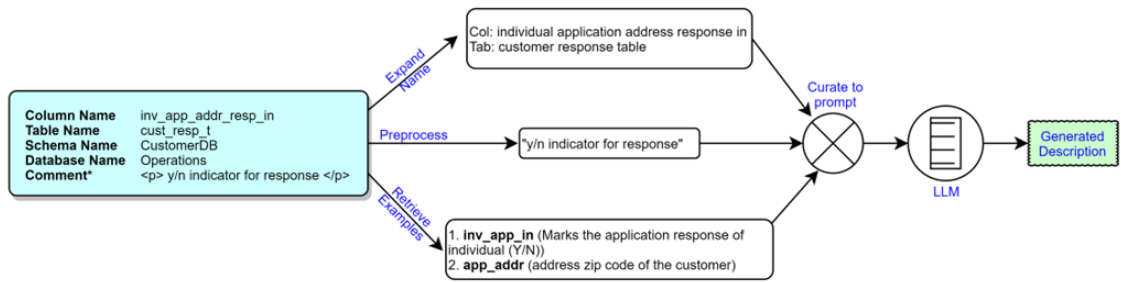

Figure 1: Column description generation framework

## 2. RELATED WORK

The critical role of contextual metadata in data systems is widely recognized. Altun, Osman et al. [7] emphasized the importance of the FAIR principles (Findability, Accessibility, Interoperability, and Reusability) for data resources [8]. While their research primarily focused on research data/software, the findings are equally applicable to industrial scenarios. Their methodology involved using existing metadata from csv/xlsx files for data annotation.

In industrial contexts, as highlighted by Eberhard Hechler et al. [9], just making data accessible is not sufficient, there is also a need for discoverable, understandable and near real time consumable data to get relevant insights out of it. A major challenge arises from contextual metadata being often not directly available. Column Type Annotation (CTA), which involves annotating the columns of a relational table with the semantic type of the values in each column, is typically the initial step for metadata enrichment. Several studies have addressed this task by aligning table columns with properties of a knowledge graph [10] or by employing transformer-based language models with fine-tuning [11]. Keti Korini et al. [12] enhanced the CTA task by using specific instructions and a two-step pipeline with ChatGPT, achieving F1 scores of over 85%. However, CTA relies on the values of the table columns, which are often access-restricted in large enterprises.

In the context of data catalog applications, it is necessary to create metadata based on the available information, which is primarily other metadata and a few records of manually annotated data. Fede Nolasco [13] discussed the use of generative AI for creating descriptions, guiding Large Language Models (LLMs) using technical/business context. However, this approach depends on an initial human input draft and sample data, which may not always be accessible. Nomadiclabs [14] demonstrated a simple method of prompting a Gemini model with available column information to generate a description. While these studies provide a good starting point, they are preliminary and limited to generic datasets with trivial columns, which differ significantly from those typically encountered in industrial settings.

Teruaki Hayashi et al. [15] in their work have shown promising results with using LLMs for data exploration and discovery. The work however focuses more on discoverability aspects using similarity searches and does not cover enriching the missing metadata content that could further aid the process. Elita Lobo et al. [16] focused on mapping column metadata to business glossaries using multi-shot in-context learning /Classification or Multiple-Choice Q&A. They relied on the ability of LLMs to generate content based on defined context. While this study effectively utilizes the capabilities of LLMs, it does not assess the hallucinations often encountered with generative models. Sayed Hoseini et al.[17] have in their initial experiments demonstrated the applicability of customizing and optimizing LLMs for semantic labelling and modelling tasks. Our approach extends this work by using a business glossary along with existing labelled metadata descriptions. We narrow down the multi-shot examples to cover all the terms in the name and maintain context relevance by employing rule-based filters and preferences. This guides the context to a more specific and narrowed-down domain, thereby reducing the likelihood of hallucinations.

Notably, there is a lack of availability of well-curated, publicly accessible datasets. VC-SLAM [18] being one of the very few we could find in this context, however due to lack of popularity and thus reliability our study relies on a proprietary dataset from our organization. This also enables us to validate the framework with the help of data owners and experts within the organization. The applicability of the study however through minor altercations should translate to other domains as well.

*Commercial solutions analysis:* The market for data catalog solutions is an emerging one and a lot of players are looking to integrate Generative AI into their arsenal for description generation. Collibra [19] uses the Google Vertex AI [20] to assist users in adding description of assets within the Collibra platform by leveraging other metadata like asset names, location in organizational structure and schemas. The AI Autodocumentation by Dataedo [21] is another such solution leveraging OpenAI [22] models along with metadata present for the asset. Although these solutions are vital in the catalog enrichment task but lack the traceability and reliability due to black-box approach and involvement of LLMs prone to hallucinations. The work we introduce in the paper can form a solid base to scale such integrations with a focus on reliability of generations through an intelligent enrichment of information (already available asset descriptions and metadata) passed to LLMs for generation.

## 3. PROPOSED METHODOLOGY

In this research, we introduce a novel approach aimed at enhancing the generation of metadata, with a particular focus on column and table descriptions. The overall pipeline is explained in Figure 2 & 3 for column description generation and table description generation respectively. The "Column Input" and "Table Input" contain similar information as illustrated in Figure 1. We employ strategies for enrichment of available information under the "Expander Pipeline" and "Preprocessing Text" that goes into the prompt for the language model. The details of which are covered in Section 3.1. Building upon this with "Similar Examples Retrieval", we further augment the prompt with examples extracted from pre-existing, carefully curated metadata and business glossaries housed within our database. This process is thoroughly explained in Section 3.2. The table description generation framework in Figure 3 utilizes the "Column Description Generation" which is covered in Figure 2. Post which a subset of columns based on relevance is taken in the "Column Selection" block to be used in the prompt curation. Finally, in Section 3.3, we provide an in-depth discussion on the large language models utilized during our experimentation. This includes a comprehensive overview of the specific prompts and instructions used.

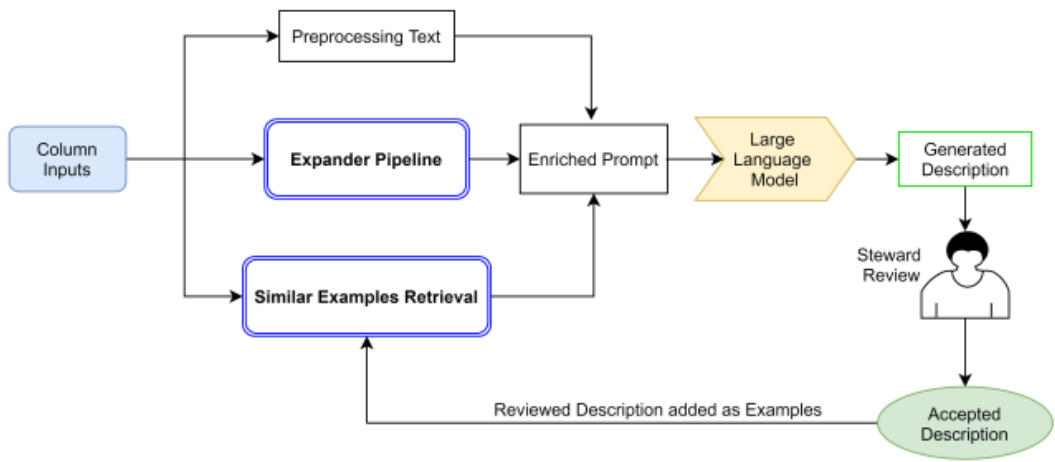

Figure 2: Column description generation framework

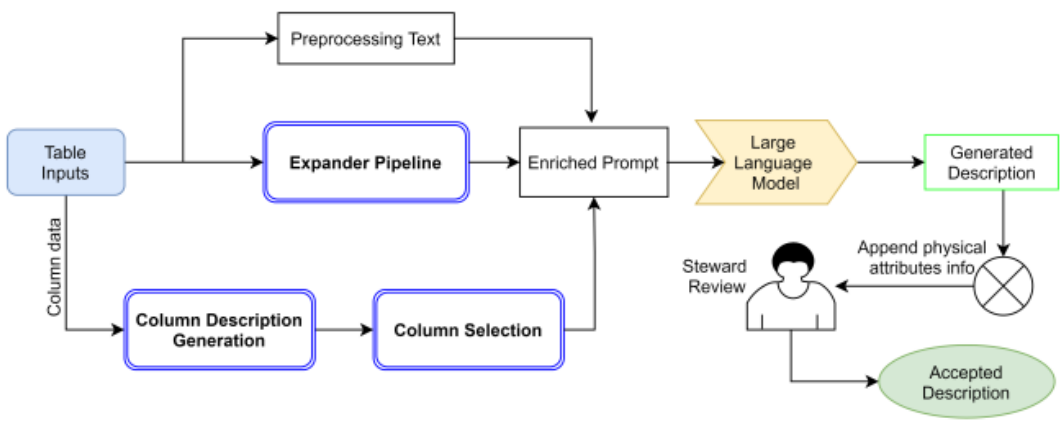

Figure 3: Table description generation framework

## 3.1. Existing metadata enrichment

The descriptions are generated based on the data inputs in the prompt. Table 1 contains the specific inputs employed in the prompts for the generation of column and table descriptions.

Table 1: Inputs considered for the study.

| Column | Column name, expanded column name*, table name, data source name, column comment if available, similar columns as few-shot examples |
|---|---|
| Table | Table name, table name expanded*, Important columns and their descriptions, business context, table comment if any |

In the initial preprocessing phase, any available comments or textual context are systematically processed. This involves removing HTML tags, eliminating multiple whitespaces, and implementing standard text cleaning procedures. Column and table names often comprise abbreviated terms. Without proper expansion of these terms, the language model attempts to expand them using its worldly knowledge, which frequently introduces high variability in the generation and results in hallucinations. To address this, the expander pipeline is employed. This pipeline leverages a meticulously crafted mapping of abbreviations to their corresponding full forms, which are then utilized to generate the expanded names. In instances where multiple expansions are possible, the pipeline incorporates a disambiguation methodology. The specifics of this methodology are elucidated in Figure 4. In cases where the disambiguation is inconclusive, the word is not expanded. Instead, it is left to the discretion of the Language Model (LLM) to interpret it using the remaining information from the prompt.

Few examples of expansions using expander pipeline are as follows.
    ytd_dist_amt        year to date distribution amount
    opt_ctrt_exp_d      option contract expiration date.
    shr_exp_d           share expiration date

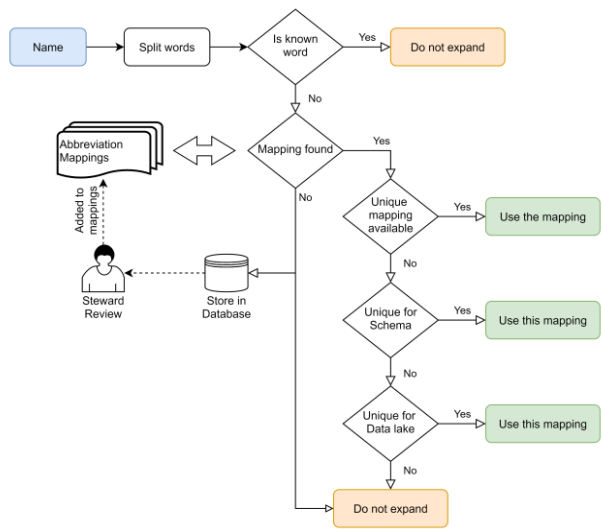

Figure 4: Expander Pipeline, disambiguation methodology

## 3.2. Retrieval Pipeline for Few-shots

Within any enterprise setup, multiple stewards curate the data catalog, each responsible for their respective data sources or schemas. The data catalog contained more than 100K existing columns, each with annotated descriptions (either source comments or descriptions).

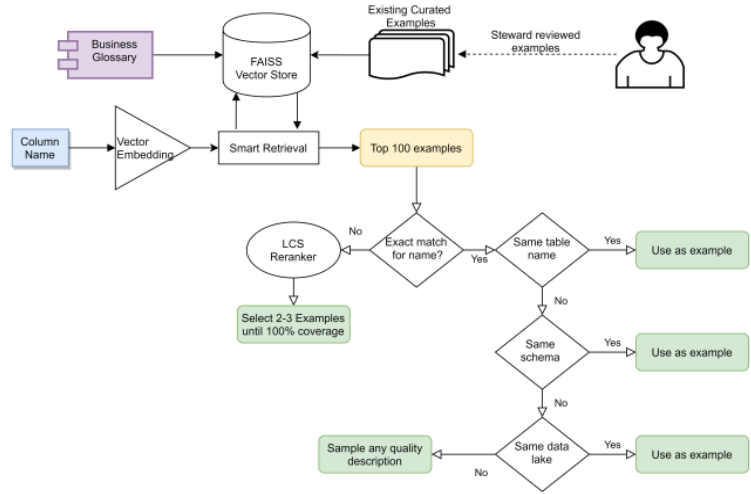

Figure 5: Similar Example Retrieval process diagram

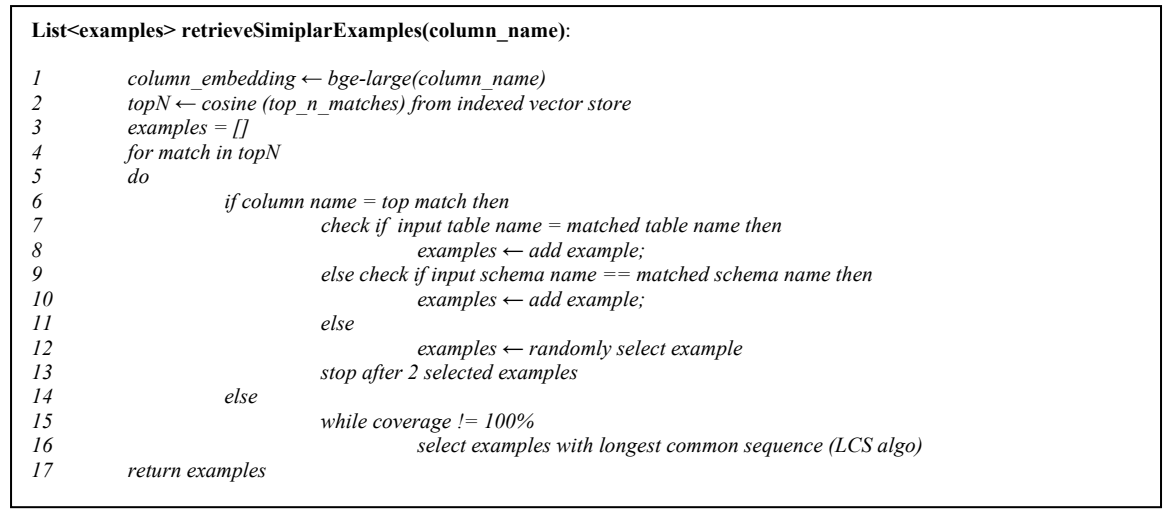

Figure 6: Pseudocode for Similar Example Retrieval process

Given that a similar column can exist in different tables across various sources, the framework leveraged these descriptions as few-shot examples for our model. We hypothesize that better semantically similar examples will provide more useful information than randomly selected examples. All existing columns were indexed in the FAISS vector library employing an embedding model (BAAI/bge-large-en-v1.5) for creating vectors for column names [23][24]. The retrieval of similar columns was a two-step process involving re-ranking:

1. Vector Similarity: The top 100 examples were retrieved using L2-normalized vector similarity with a flat index. In instances where an exact match was found, preference was given to examples belonging to the same table name or data source name over matching examples from different tables or data sources.

2. Re-ranking: In situations where no exact match was found, the top 2-3 examples were selected from the top 100 examples retrieved in step 1. This selection was made using a Longest

Common Sub-sequence (LCS) matching algorithm. The examples were chosen to ensure that all words in the column name were fully covered using the retrieved examples (a maximum of 3). An illustration of this process is provided in Figure 5 and 6.

## 3.3. Generative Large Language Models

In the study, we conduct a comparative analysis of various Large Language Models (LLMs). For column descriptions, we examine three models namely pretrained Llama2-13b-chat, Fine-tuned Llama2-7b-chat, and open AI GPT3.5 turbo model. For table descriptions, our comparison includes pretrained Llama2-13b-chat, Open AI GPT 3.5 Turbo and GPT 4 models [25][26][27]. The selection of language models was primarily influenced by the resources required for deployment and the associated costs.

### 3.3.1. Fine-tuning models

Generating column description is a short text generation task. Hence, we trained a Llama2-7b model over a curated dataset comprising over 35,000 example pairs of column information and corresponding curated descriptions or business glossary. Since we were constrained to use a reasonably small GPU instance the training was performed using QLoRA[28] technique. In QLoRA the number of trainable parameters are reduced to a fraction of total parameters, this is accomplished by a rank decomposition of the weight update matrix into two lower rank matrices as proposed in LoRA[29]. Also the model weights are quantized into lower precision thus helping in loading LLMs in GPU restricted environments. The specifics of the fine-tuning process are detailed in the Appendix A.1.

### 3.3.2. Few-shot prompting on pretrained models

We have used GPT 3.5 Turbo as the model for evaluation, we did not evaluate GPT 4 for column descriptions due to its comparatively higher cost, which could pose a bottleneck in an enterprise setup with a large volume of columns. The structure of the prompt used for the model is depicted in Figure 12 (Appendix A.3). For columns, we employ processed inputs, expanded names, business glossary, retrieved examples, and system instructions to guide the model on how to process the information and generate the desired output.
Table descriptions, being typically lengthy and lacking a standard template, necessitate a different approach. We curate the prompt to pose directed questions to the language model and stitch the responses together to form the description. This process utilizes processed inputs, a selection of columns (approximately 25), additional business context or user-specified context, and a set of instructions. The process of column selection is executed following a logical hierarchy, detailed as follows: Initially, audit columns are eliminated. Subsequently, all significant columns, as identified in the data catalog are selected. The next step involves choosing the first five columns, which are the primary keys. This is followed by the selection of the most highly ranked columns, based on user popularity. Finally, a random sampling of the remaining columns, which contain rich information, is performed.

The generated description is further supplemented with physical information about the table, such as the data source and update frequency, where available. The outputs generated by the model undergo post-processing, which includes JSON processing and domain-specific corrections accumulated over time. The final processed generation is then forwarded to the stewards for review before being updated in the database.

# 4. EXPERIMENTS AND RESULTS

## 4.1. Dataset

The study has been performed on proprietary data within the organization due to lack of well curated publicly available sources. The data from within the organization was also preferred as it was easier to validate the findings from the study with the domain experts available within the organization. The models' performance evaluation was observed on two datasets for table and columns description generation.

1. Existing Ground Truth Dataset: This is a curated validation set of examples for 8088 columns & 122 tables. The validations set is created by sampling examples from an existing dataset of well-defined descriptions considering examples across different data lakes and schemas.
2. Data Steward's Feedback Dataset: This dataset consists of 892 columns and 31 tables with the model generated descriptions. The stewards were asked to curate it by making edits to it.

## 4.2. Performance Evaluation

*Existing Ground Truth Dataset*: We calculate Bert-score [30] to analyze the performance evaluation among models between generated text and existing ground truth. Table 2 and Table 3 shows bert-score for generated tables and columns descriptions respectively in the ground truth.

Table 2: Ground Truth Data: Table Description

| Metrics | Evaluation: Bert Score | | |
|---|---|---|---|
| | Llama2-13B | GPT3.5 turbo | GPT-4 |
| Mean Precision | 0.49 | 0.45 | 0.46 |
| Mean Recall | 0.53 | 0.53 | 0.55 |
| Mean F1 | 0.50 | 0.48 | 0.50 |

Table description generation is a long-form text generation NLP task. For such tasks, Bert-score may not be the right metric as one can observe low Bert-scores from models with marginal random difference. The scores are expected to be low as data stewards have high business and domain specific knowledge not available to model. The generated descriptions should only be considered as jumpstart for data curation in case of tables. Structurally, the experiments and manual reviews establish that GPT-4 follows instructions appropriately and generates more consistent text description structurally whereas Llama, GPT 3.5 had inconsistent structures.

Table 3: Ground Truth Data: Column Description

| Metrics | Column Description: Bert Score | | |
|---|---|---|---|
| | Llama2-13B | GPT3.5 turbo | Fine-Tuned Llama2-7B |
| Mean Precision | 0.66 | 0.70 | 0.75 |
| Mean Recall | 0.68 | 0.65 | 0.73 |
| Mean F1 | 0.66 | 0.67 | 0.74 |

Overall, the fine-tuned llama-2 model scores highest for column description generation. However, further close examination of generated column descriptions based on the information provided in the prompt (column comment and relevant examples), it was found that GPT Turbo

3.5 performed better than fine-tuned model. The below table 4. shows the examining Rouge-1 score [31] with example descriptions or comments. Copied column represents the count of instances where Rouge-1 score between the example/comment description and the generation was greater than 0.95). This ensures that models exactly (lexical, not semantic) copies information from inputs in prompt.

Table 4: Evaluation: Copying Behavior

| Models | Exact Matched Example Scenario (Columns) | | |
|---|---|---|---|
| | Total Instances | Copied | %Copied |
| Fine-tuned Llama-2 7B | 7259 | 3617 | 49.47% |
| GPT 3.5 turbo | 7259 | 575 | 7.1% |

It is evident that GPT model tries to recompose all information in the prompt and paraphrase it better, whereas Finetuned Llama model does not use all information present but copy and pastes text from column comment or exact matches from few-shot examples.

*Data Steward's Feedback Dataset*: Rouge-1 score provides a comprehensive indication of editing done while final curation of descriptions unlike Bert-score which provides semantic similarity [30][31]. The data stewards were also asked to provide a label depending on minor or major editing required in the curation process. Table 5 shows encouraging results leading to saving of manual effort of content creation.

Table 5: GPT3.5 Turbo & GPT-4 Performance

| Categories / Metrics | Column Descriptions | Table Descriptions |
|---|---|---|
| Acceptable as is or with minor edit | 31 (87%) | 786 (88%) |
| Acceptable with major edit | 4 (12%) | 106 (12%) |
| Rouge-1 F1 score (mean) | 0.81 | 0.87 |
| Bert-score F1 (mean) | 0.89 | 0.91 |

The high mean Rouge-1 F1 scores are also indicative of minimal efforts required in modifications by the stewards when provided with generated description as jumpstart.

## 4.3. Factual Grounding

Yuheng Zha et. al. proposed AlignScore, a factual consistency metric based on a unified text-to-text information alignment function to evaluate model generations by comparing it with the information provided in the prompt [32]. This classification score is indicative of whether the generation is entailing, contradicting or neutral with the provided information in the prompt.
There were majorly three inputs provided in the prompt namely few-shot examples, source comment and metadata related to columns (expanded column name, table name and data lake etc.). The following figures (Figure 7-10) show a comparison of AlignScore for the three models in column description generation scenarios for a detailed study in cases:
1. When there are exact matches or only partial matches in the retrieved examples (Exact Example/ No Exact Example).
2. When the source comments are missing or present for the column (Comment Missing/Comment Present).

The x-axis represents Ground Truth Alignment, i.e. how well the ground truth is entailed by the inputs provided in the prompt. We study the 4th quartile where ground truth also entailed information in the prompt using same alignment model. The Count distribution sub-figure shows the number of samples in each quartile for the Ground Truth Alignment values.

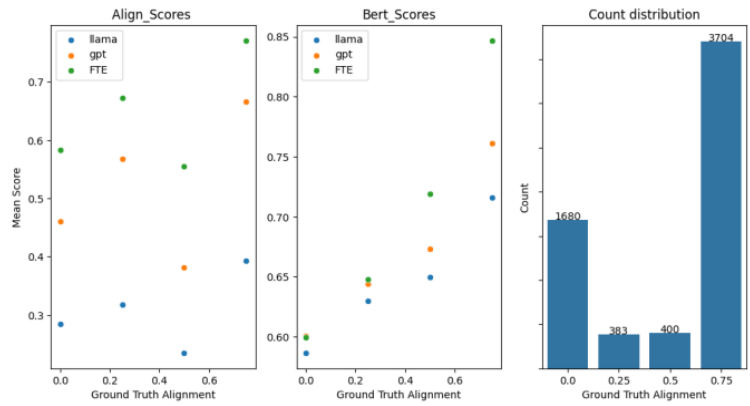

Figure 7. AlignScore across quartiles (scenario 1)

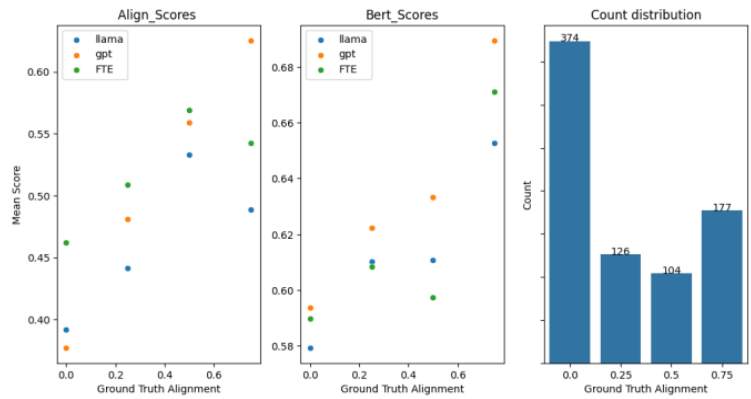

Figure 8. AlignScore across quartiles (scenario 2).

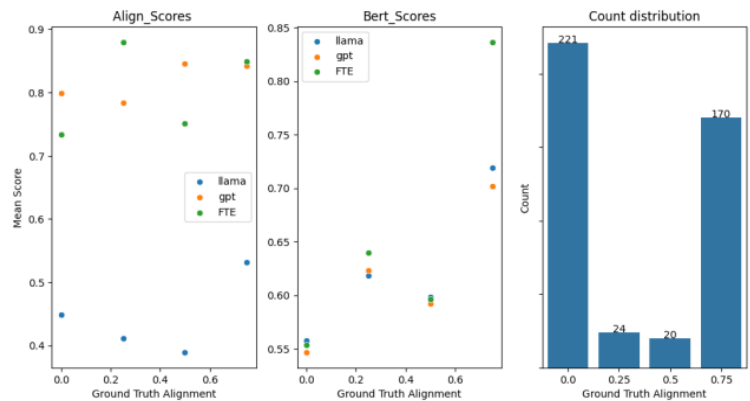

Figure 9. AlignScore across quartiles (scenario 3).

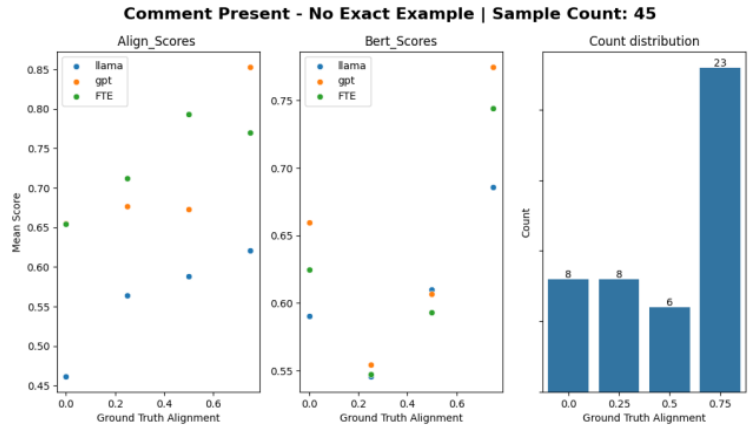
Figure 10. AlignScore across quartiles (scenario 4).

Bert-scores for respective quartiles are demonstrated for relative comparison. Following were the observations:
    a. Both fine-tuned Llama2-7B and GPT 3.5 model entailed prompt inputs with high AlignScore (>0.65) when exact matched columns were retrieved. Comparatively, the finetuned model performs marginally better than GPT3.5 turbo due to copying by former and paraphrasing by latter.
    b. Whenever there is partially matched example found, GPT models performs better than Finetuned model as it was able to apprehend the column names in context of a database and generate description.
    c. Pretrained Llama2-13B distinctly gets ruled out due to consistently lower factual grounding score inferring lesser entailment ability. On the contrary, Bert-score could not demonstrate the same due to high semantic similarity scores between financial domain specific texts.

## CONCLUSIONS

Any pretrained model without provided domain knowledge may lead to ambiguous columns and table description generation specially for abbreviated term definition. In this paper, we propose a smart data catalog, an LLM based framework with few-shot retrieval and additional existing attributes to generate descriptions. The research conducted extensive experiments with prompt enrichment (bringing domain knowledge) to propose unique components (few-shot retrieval, expander) which can assist LLMs in generating more accurate descriptions. These components are developed leveraging the idea of existing curated metadata content for large data catalogs. We also perform deeper analysis on the behavior of implemented LLM models and highlight the potential concern of finetuned model of copying prompt text in description several times. On the contrary, few-shot enriched prompting with state-of-the-art GPT models paraphrased and entailed the prompt inputs accurately. The paper pens evaluation metrics which includes semantic accuracy and factual alignment with existing ground truth and acceptance accuracy with post-generation curated descriptions (Rouge-1).

**Future work**. As future directions, we wish to explore deeply the techniques for in-domain knowledge adaptation including better finetuning or retrieval augmented generation approaches. With the advent of newer LLMs and ongoing research, we intend to explore recent models which may further enhance the generation performance. The advent of LLMs with significantly larger context windows also makes room for use of supplementary knowledge articles that can provide additional context for generations. In scenarios where the underlying data is accessible and does not pose any shareability concerns, the enrichment of prompt could also involve using a sample of actual values for better alignment. Additionally, although we use a methodology for filtering the most relevant context that goes to the prompt per generation, there is still room for improvement in the use of data model for the task. The availability of data models as inputs to language models can further help them judge the relationships among assets in a better way.

ETHICS STATEMENT

The research proposes an LLM based framework to automate the metadata content generation to be used for data catalogs. While it has shown encouraging results, it can produce factually incorrect results. LLM experimental results and behavior analysis penned in the paper may not be appropriate for other applications or non-identical data catalog scenarios.

*Disclaimer*: The view or opinion expressed in this paper are solely those of the author and do not necessarily represent those of Fidelity Investments. This research does not reflect the procedure, processes, or policies of operations within Fidelity Investments.

# A. APPENDIX

In the appendix section, we provide additional details about model finetuning, toxicity detection and prompt illustrations which could not be included in the main paper.

## A.1. Fine-tuning Llama2-7B model

We trained the finetuned model on a total of 35K+ pairs of prompt and response were chosen after EDA analysis of 500K column names and its meta information. These pairs included only pairs where curated column description or business term definition were available. All training pairs were under 1024 token length. We employ QLORA technique to finetune Llama2-7b-chat model using 'causal_lm' task with train and validation split of 90:10. We observe that increasing the number of steps (epochs) resulted in overfitting and absurd results. Hence, the training process was limited to 1 epoch. Figure 11. shows the training and validation loss over 1K steps.

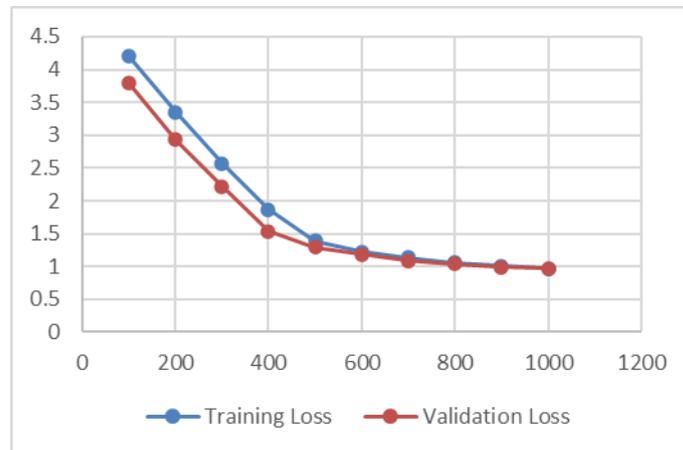

Figure 11. Training and validation loss over 1K steps

## A.2. Toxicity Evaluation

With respect to guardrails, we employed toxicity detection utilizing open-source library 'detoxify' for all generated text by the proposed framework [33]. We can observe that the toxicity scores are too low for such LLM applications in metadata management use-case as seen in Table 6 and Table 7. Additionally, the GPT models captures toxic language generations for (hate/self-harm/sexual/violence) and none of the example generations were flagged.

Table 6: Toxicity Score: GPT3.5 Turbo (Column Description)

| Categories / Metrics | Ground Truth | GPT 3.5 Turbo |
|---|---|---|
| Toxicity | 0.00077 | 0.00068 |
| Severe_Toxicity | 0.00012 | 0.00012 |
| Obscene | 0.00022 | 0.00019 |
| Threat | 0.00012 | 0.00012 |
| Insult | 0.00019 | 0.00019 |
| Identity_attack | 0.00014 | 0.00015 |

Table 7: Toxicity Evaluation: GPT-4 (Table Description)

| Categories / Metrics | Ground Truth | GPT 4 |
|---|---|---|
| Toxicity | 0.00066 | 0.00060 |
| Severe_Toxicity | 0.00012 | 0.00013 |
| Obscene | 0.00019 | 0.00020 |
| Threat | 0.00013 | 0.00013 |
| Insult | 0.00018 | 0.00018 |
| Identity_attack | 0.00014 | 0.00014 |

## A.3. Prompt Illustration

Figure 12. below shows an input column name and its additional information which is utilized to fetch meaningful inputs (expanded names and semantically similar examples) to further enrich the prompt template. The dynamically retrieved examples were passed in as a list of messages between the user prompt and assistant response instead of long concatenated string.

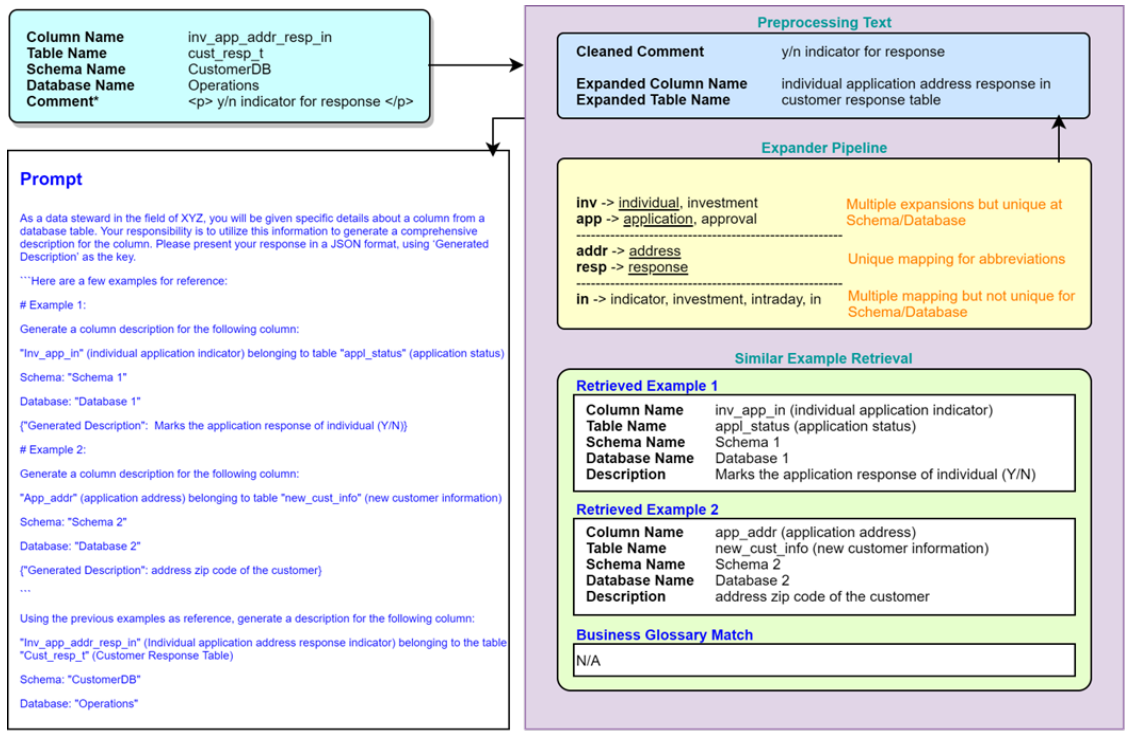

Figure 12. Illustration of Prompt Enrichment.